\documentclass[conference]{IEEEtran}
\IEEEoverridecommandlockouts
\usepackage{cite}
\usepackage{amsmath,amssymb,amsfonts}
\usepackage{algorithm}
\usepackage{algorithm}
\usepackage{algpseudocode}
\usepackage{graphicx}
\usepackage{textcomp}
\usepackage{xcolor}
\usepackage{bm}
\usepackage{multicol}
\usepackage{listings}
\usepackage{bbm}
\usepackage{url}
\usepackage{mathtools}
\usepackage{dashrule}
\usepackage[left=1.02in,right=1.02in,top=0.75in, bottom=1.04in]{geometry}
\algnewcommand{\IIf}[1]{\State\algorithmicif\ #1\ \algorithmicthen}
\algnewcommand{\ElseIIf}[1]{\algorithmicelse\ #1} 
\algnewcommand{\EndIIf}{\unskip\ \algorithmicend\ \algorithmicif}
\def\BibTeX{{\rm B\kern-.05em{\sc i\kern-.025em b}\kern-.08em
    T\kern-.1667em\lower.7ex\hbox{E}\kern-.125emX}}
\begin{document}

\newtheorem{example}{Example}

\title{A Fast Quantum Image Compression Algorithm based on Taylor Expansion\\

\thanks{}
}

\author{
	\IEEEauthorblockN{Vu Tuan Hai\textsuperscript{1}, Huynh Ho Thi Mong Trinh\textsuperscript{2,3}, Pham Hoai Luan \textsuperscript{2,3}}
	\IEEEauthorblockA{
    \textsuperscript{1} Nara Institute of Science and Technology, 8916–5 Takayama-cho, Ikoma, Nara 630-0192, Japan.\\
    \textsuperscript{2} Faculty of Software Engineering, University of Information Technology, Vietnam.\\
    \textsuperscript{3} Vietnam National University, Ho Chi Minh City, Vietnam.\\
Email: vu.tuan\_hai.vr7@naist.ac.jp} 
}

\maketitle

\begin{abstract}

With the increasing demand for storing images, traditional image compression methods face challenges in balancing the compressed size and image quality. However, the hybrid quantum-classical model can recover this weakness by using the advantage of qubits. In this study, we upgrade a quantum image compression algorithm within parameterized quantum circuits. Our approach encodes image data as unitary operator parameters and applies the quantum compilation algorithm to emulate the encryption process. By utilizing first-order Taylor expansion, we significantly reduce both the computational cost and loss, better than the previous version. Experimental results on benchmark images, including Lenna and Cameraman, show that our method achieves up to 86\% reduction in the number of iterations while maintaining a lower compression loss, better for high-resolution images. The results confirm that the proposed algorithm provides an efficient and scalable image compression mechanism, making it a promising candidate for future image processing applications.

\end{abstract}


\section{Introduction}

With the increasing development of applications that primarily store image data, such as message applications, the demand for exchanging images has grown rapidly. Given the limitation of memory capacity and channel bandwidth, image compression is essential for optimizing hardware resources. Image compression remains a fundamental challenge with broad implications; the objective is to reduce image size by minimizing redundancy and irrelevant information. Early compression techniques, such as JPEG and JPEG2000, employed discrete cosine transform \cite{john2021discrete}, wavelet transforms \cite{Shensa1992TheDW}, and entropy coding \cite{10.1007/978-3-030-58520-4_27}. Recent advancements leverage techniques such as principal component analysis (PCA) \cite{10.1007/11596356_90} and neural networks \cite{8464188}, including Convolutional Neural Networks (CNNs) \cite{Mentzer_2019_CVPR}, Autoencoders (AEs) \cite{9277919}, and Recurrent Neural Networks (RNNs) \cite{Toderici_2017_CVPR}. CNN-based approaches have demonstrated superior performance over traditional methods by learning hierarchical image features through deep networks and end-to-end encoder-decoder architectures. AEs and RNNs, which effectively reduce dimensionality and generate compact image representations, further enhance compression efficiency, particularly at lower bit rates, where conventional methods often struggle with artifacts. Despite advancements in image compression, all existing methods face the trade-off between file size and image quality. For instance, while JPEG provides the best compression levels with lossy encoding - leading to quality degradation upon repeated saving - PNG ensures lossless property but results in larger file sizes, making it unsuitable for high-resolution images. Given these limitations, no single compression method is universally optimal.  

To address this challenge, our research explores a machine learning-based approach, introducing a hybrid quantum-classical model \cite{schuld2018supervised} that redefines the conventional paradigm of image compression. This approach leverages quantum computing principles, specifically compilation algorithms \cite{Khatri2019quantumassisted}, to encode an image as rotation angles. The size of angles only grows logarithmically based on the image size, achieving compression efficiency. We focus on lossy compression, which refers to techniques that employ approximate methods. This contrasts with lossless data compression, which maintains data integrity. Our method is named Fast Quantum Image Compressor (Fast QIC) which is an improved version of Quantum Image Compressor (QIC) \cite{hai_encrypted} but with optimized resources and smaller loss. The baseline QIC with related background techniques is mentioned in Section~\ref{sec:background}. Section~\ref{sec:approach} presents our proposed Fast QIC. Section \ref{sec:experiments} discusses the experimental results on two basic images. Finally, Section \ref{sec:conclusion} concludes the paper.

\section{Background}
\label{sec:background}

\begin{figure}
    \centering
    \includegraphics[width=0.99\linewidth]{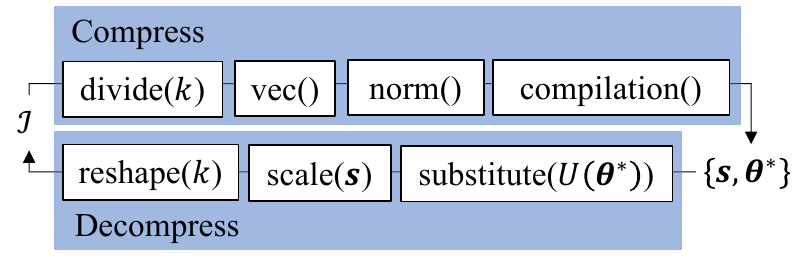}
    \caption{Overview of quantum image compression/decompression procedures.}
    \label{fig:introduce}
\end{figure}

\begin{figure*}
    \centering
    \includegraphics[width=0.99\linewidth]{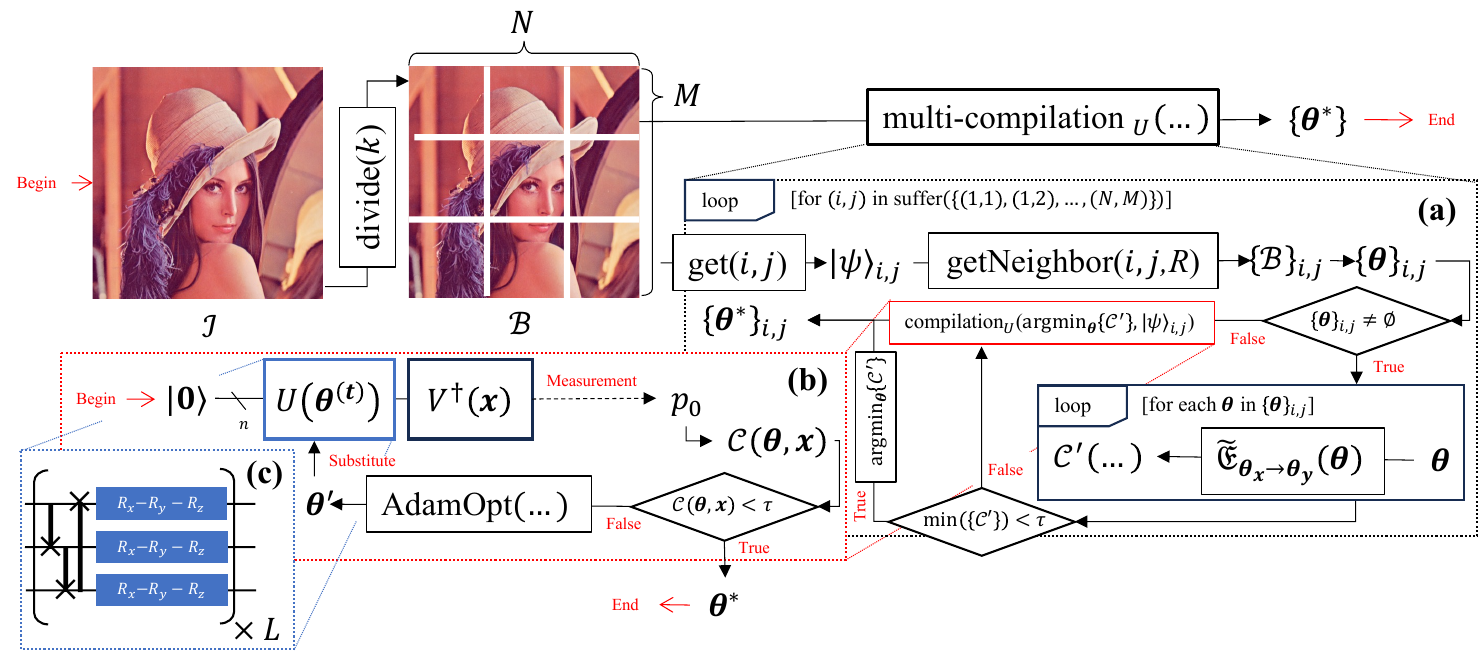}
    \vspace{-0.4cm}
    \caption{The architecture of fast QIC, $\text{get}(i,j)$ take the blocks $i,j$, then apply $\text{vec}\;\circ\;\text{norm}\;(\ldots)$ to return corresponding quantum state $|\psi\rangle_{i,j}$ (a) Looping through all blocks generated from the image $\mathcal{I}$ (b) The compilation algorithm inside each loop (c) A 3-qubit $W_{\text{chain}}+XYZ$ ansatz as $U(\bm\theta^{(t)})$.}
    \label{fig:overview}
\end{figure*}

\subsection{Quantum Image Compressor}

The quantum image compression problem relates to the image encoding in the quantum circuit. An image is 
presented as the color density and position for each pixel. The general depiction of a quantum image is outlined as $|\mathcal{I}\rangle = \frac{1}{\sqrt{2^n}} \sum_{j=0}^{2^n-1} |C_j\rangle \otimes |j\rangle$, where $|C_j\rangle$ denotes the color data and $|j\rangle$ encodes the position. There are three main methods, including FRQI \cite{Sang2016}, NEQR \cite{Wang2022}, and QHSL \cite{YAN2021196}. FRQI and NEQR differ in their approaches to quantum image encoding. FRQI leverages superposition to minimize the number of qubits needed for encoding but requires more complex operations during measurement. In contrast, NEQR employs basis encoding, resulting in lower computational complexity and a broader range of transformations while also achieving more precise image recovery. Although both FRQI and NEQR are designed for grayscale images, they can be extended to color images by applying the grayscale encoding three times for the RGB channels. Meanwhile, QHSL is based on the HSL model, offering a color encoding technique that aligns more closely with human perception. Unlike FRQI and NEQR, QHSL encodes color information by integrating amplitude values and an additional qubit sequence. For QIC, this method separates an image as blocks; then there is no need to encode the position. The color density will be encoded as quantum amplitudes directly since $|\psi\rangle\langle\psi|=1$, it is ranged in $[0,1]$ instead of $[0,256]$. 

The QIC's compress/decompress procedure is presented in Fig.~\ref{fig:introduce}. We receive the ($\mathcal{I}_N\times \mathcal{I}_M$) image $\mathcal{I}$ and block size $k$. The image $\mathcal{I}$ is divided into $k\times k$ - blocks, denoted as $\{\mathcal{B}_{1,1}, \mathcal{B}_{1, 2}, \ldots,\mathcal{B}_{N,M}\}$. In case $\mathcal{I}_N /\mathcal{I}_M\bmod k \neq 0$, we add the padding zero with the padding size:

\begin{align}
    &P_N = k - (\mathcal{I}_N \bmod k),\nonumber\\
    &P_M=k-(\mathcal{I}_M\bmod k);
\end{align}

then $N=(\mathcal{I}_N+P_N)/k$ and $M=(\mathcal{I}_M+P_M)/k$ is the number of blocks by row and column.  Each block is vectorized and normalized as a quantum state (length is 1) by \verb|vec()| and \verb|norm()| functions, respectively. The normalized factor $s_{i,j}$ is saved for reconstruction and the quantum state is prepared by the compilation algorithm, which is presented in Section.~\ref{sec:compilation}.  Any $n$-qubit state $|\psi\rangle=U(\bm\theta)|0\rangle^{\otimes n}$ can be expressed as a learnable operator $U$ and its parameter vector $\bm\theta$, where  $n=2\lceil\log_{2}k\rceil$ qubits and $|\bm\theta|=\mathcal{O}(\text{poly}(n))\ll k^2$, this is the quantum advantage in our previous quantum-classical image encryption scheme \cite{hai_encrypted}. Notice that $\theta_j\in[0,2\pi]\;\forall\theta_j\in\bm\theta$. Since $k,n\in \mathbb{R}^{+}$, $k$ should be in the form of $2^z$ to avoid zero padding. In the decompress process, we take $|\psi\rangle$ by substituting $\bm\theta$ back to $U$, then convert $s|\psi\rangle$ to block $\mathcal{B}_{i,j}$ via function $\verb|reshape|(k)$. Finally, the original image is reconstructed by arranging blocks from left to right and from top to bottom.

\subsection{Quantum compilation algorithm}
\label{sec:compilation}

Parameterized Quantum Circuits (PQCs) are a fundamental framework in quantum machine learning \cite{quantumcircuitlearning, PhysRevLett.122.040504} or hybrid quantum-classical models which consist of fixed and trainable gates. These parameters can be optimized by the Parameter-Shift Rule (PSR) technique through $2m$ quantum evaluations \cite{Wierichs2022generalparameter}.  A popular algorithm - quantum compilation \cite{Khatri2019quantumassisted} which applies to many problems such as quantum state preparation and quantum tomography \cite{Hai2023tomography} uses PQCs as a basic component. This algorithm transforms an initial state into a desired target state, as shown in Fig.~\ref{fig:overview}-(b). In the context of image compression, quantum states can be replaced with vectorized blocks, allowing the compilation algorithm to transform an image into the parameters. 

We start with the $n$ - qubit zero state $|\psi^{(0)}\rangle=U(\bm\theta^{(0)})|0\rangle^{\otimes n}$
under a ($L$ layers) trainable unitary $U\left(\bm{\theta}\right)=\bigotimes_{j=0}\left(U_j\left(\bm{\theta}_k\right) \otimes V_j\right)$ where $\bm\theta^{(0)}=\mathbbm{1}\equiv[1\;1\;\ldots\;1]^{\intercal}$. $U_j$ and $V_j$ are variational and fixed parts, respectively. The target state is notated as $|\psi\rangle$ and the closeness of the two states is given by the inner product $ d(\psi^{(t)},\psi)=|\langle\psi^{(t)}| \psi\rangle|^2$. The target state $|\psi\rangle$ exists ($|\psi^{(t)}\rangle\approx|\psi\rangle$) when its distance reaches $0$. Because $d(.,.)=1$ mean $|\psi^{(t)}\rangle$ and $|\psi\rangle$ are overlapped, thus, we minimize:

\begin{align}
\bm {\theta}^*=\underset{\bm {\theta}}{\operatorname{argmin}} (1-d(\psi^{(t)},\psi)),
\label{eq:optimize}
\end{align}

noticed that $\mathcal{C}^*\equiv\mathcal{C}(\bm\theta^*)=1-d(\psi^{(n_{\text{iter}})}, \psi)$ is the minimal cost/loss value. The Equation~\eqref{eq:optimize} can be achieved by some first-order optimizers, such as Adam \cite{kingma2014adam}, which requires the gradient $\nabla_{\bm\theta}C(\bm\theta)=\left[\partial_{\theta_0}\mathcal{C}(\bm\theta)\;\partial_{\theta_1}\mathcal{C}(\bm\theta)\;\ldots\;\partial_{\theta_{m-1}}\mathcal{C}(\bm\theta)\right]^{\intercal}$.

\begin{algorithm}[ht]
\caption{$\text{compilation}_U( \bm\theta^{(0)}, x)$} 
\label{algo:pstabilizer}
\begin{algorithmic}[]
\Require $\bm\theta^{(0)},x$

\For{$t$ in $[0,1,\ldots,n_{\text{iter}}-1]$}
    \State $p_0\gets\mathcal{M}_{\mathbb{Z}_0}(|\langle \bm 0|V^{\dagger}(x)U(\bm\theta^{(t)})|\bm 0\rangle|^2)$
    \If{$\mathcal{C}_x(\bm\theta^{(t)})<\tau$}
        \For{$j$ in $[0,1,\ldots,m-1]$}
            \State $\partial_{j}\mathcal{C}_x(\bm\theta^{(t)})\gets\text{PSR}(\mathcal{C}_x,\bm\theta^{(t)},j)$
        \EndFor
        \State $\bm\theta^{(t+1)}\gets \text{AdamOpt}(\nabla_{\bm\theta}\mathcal{C}_{x}(\bm\theta^{(t)}),\ldots)$ 
    \Else
        \State\Return $\bm\theta^{(t)}$
    \EndIf
\EndFor
\State\Return $\bm\theta^*\equiv\bm\theta^{(n_{\text{iter}})}$
\end{algorithmic}
\end{algorithm}


As shown in Fig.~\ref{fig:overview}-(b), the compilation algorithm runs on two computers: classical and quantum. The task $d(,.,)$ is executed on the quantum computer, and other tasks are performed on the classical computer. Therefore, the quantum compilation process is a hybrid quantum-classical model \cite{PhysRevLett.122.040504}. Finally, compressing any images $\mathcal{I}$ takes only logarithm scale because we only use $n=2\log_{2}N$ qubits and $\mathcal{O}(L\text{poly}(n))$ parameters; this is the quantum advantage in our proposed scheme.

\subsection{Size of compressed image}

The total number of parameters in ($L$ layer) $U(\bm\theta)$, is given by $m=m'nL$, where $m'$ is dependent on the chosen ansatz. The total number of parameters is $N M(m'nL + 1)$, where $1$ corresponds to the scale factor $s_i$. To ensure that the compressed image size remains smaller than the original, we impose the following inequality $N M (m'nL + 1) < \mathcal{I}_N\mathcal{I}_M$. Assume that we process on a block-aligned image ($\mathcal{I}_N=\mathcal{I}_M$ and $\mathcal{I}_N \bmod k = 0$) and $L$ increase by $n$, then:

\begin{align}
    &N^2(m'nL + 1)<\mathcal{I}_N^2 k ^2\nonumber\\
    \Leftrightarrow \;&k>\sqrt{4r(\lceil\log_2k\rceil)^2+1}\approx2\lceil\log_2k\rceil\sqrt{m'}
\label{eq:ineq}
\end{align}

\begin{figure}
    \centering
    \includegraphics[width=0.99\linewidth]{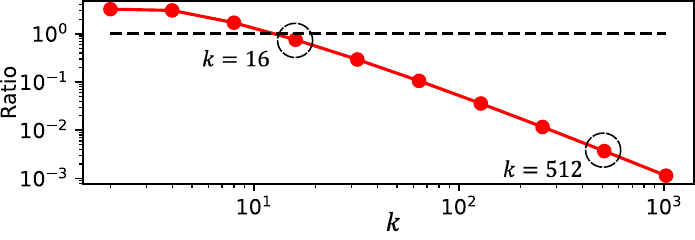}
    \caption{The ratio (data size) between compressed image and original image. The dotted line is $\text{ratio}=1$, at this line, the compressed image and the original image's size are equal. Both axes are plotted on a logarithmic scale.}
    \label{fig:ratio}
\end{figure}

This result indicates that increasing $k \in \{2,4,\ldots 2^z\}$ as much as possible leads to an exponential reduction in the compressed image size because $\sqrt{m'}$ is constant and $k$ on the right side of~\eqref{eq:ineq} appears within the logarithm function. With $W_{\text{chain}}+XYZ$ ansatz ($m'=3$) \cite{PRXQuantum.2.040309}, the inequality ~\eqref{eq:ineq} is true at $k\geq 16$. The Fig.~\ref{fig:ratio} shows the $\text{ratio}=\text{compressed image data size}/\text{original image data size}$ decrease linearly based on $k$ or exponentially based on the number of qubits $n$. 

\section{Proposed method}
\label{sec:approach}

\subsection{The fast QIC}

For each block, the compilation process is called at maximum $n_{\text{iter}}$ iterations;  each iteration refers to $2m+1$ quantum evaluations for computing $\nabla_{\bm{\theta}}\mathcal{C}(\bm\theta)$ and $\mathcal{C}(\bm\theta)$. Therefore, the naive QIC \cite{hai_encrypted} offers $N M n_{\text{iter}} (2m+1)$ quantum evaluation for an image, which increases quadratically by image size. Since the resource for executing $\verb|compilation|_U(\ldots)$ on big circuits (or deep circuits) is large, eliminating the redundancy evaluation is the target of this research. The difference with the naive QIC is that the fast QIC (proposed) considers the neighbor blocks at each step as Fig.~\ref{fig:overview}-(a). It can be explained as in the same local area, there are duplicated or similar blocks. If $\verb|compilation|_U(\mathbbm{1}, x)$ take $n_x$ iterations to converge, $\verb|compilation|_U(\bm\theta^*_{x}, y)$ with $y\approx x$ will only take $n_y\ll n_x$ iterations.

The number of adjacency blocks is determined by radius $R$. If $R=1$, there are maximal $8$ adjacency blocks of any block $\mathcal{B}_{i,j}$ considered, generally, there are $(R+2)^2-1$ adjacency blocks. The neighbor of blocks $\mathcal{B}_{i,j}$ is defined as the set $\{\mathcal{B}_{i-R,j-R},\mathcal{B}_{i-R,j-R+1},\ldots, \mathcal{B}_{i+R,j+R}\}$. This set can be less in case $i-R\leq0$ or $j-R\leq0$. Since the blocks in the same area are normally not much different, using the estimator:

\begin{align}
    \tilde{\mathfrak{E}}_{\bm\theta_{x} \to \bm\theta_{y}}(\bm\theta_x):\mathbb{R}^m\rightarrow\mathbb{R}^m
    \label{eq:estimator}
\end{align}

will reduce the total number of iterations. A block that has the corresponding parameter ($\bm\theta$) is marked as ``compiled''. If there are no ``compiled'' blocks around any block $\mathcal{B}_{i,j}$, $\verb|compilation|_U(\mathbbm{1},\mathcal{B}_{i,j})$ must be executed; otherwise, the best-estimated parameters from adjacency ``compiled'' blocks are taken as the procedure in Fig.~\ref{fig:overview}-(a). Because the adjacency ``compiled'' blocks are only available from top-middle to left-middle in case we consider $\mathcal{B}_{1,1}$ to $\mathcal{B}_{N, M}$ sequentially; the order will be suffered. 

\subsection{Taylor expansion on cost function}

In the following section, we use the notation $\mathcal{C}_{x}(\bm\theta^{*}_{x})\equiv\mathcal{C}(\bm\theta^*_x, x)$ as the loss function for $x$ $\bm\theta^{*}_{x}$.
Let $\bm\theta^{*}_{x}$ be such that $\mathcal{C}_{x}(\bm\theta^{*}_{x})\approx0$. When the state is perturbed to $y = x + \epsilon$ or $||y-x||\approx0$, we wish to find $\tilde{\bm\theta}^{*}_{y}$ such that $\mathcal{C}_{y}(\tilde{\bm\theta}^{*}_{y}) \approx 0$. 
Considering the first-order Taylor expansion of $\mathcal{C}$ with respect to $\bm\theta$ at $\bm\theta^{*}_{x}$ and fixed $y$:

\begin{align}
&\mathcal{C}_y(\tilde{\bm\theta}^{*}_{y}) \approx \mathcal{C}_y(\bm\theta^{*}_{x}) + \nabla_{\bm\theta} \mathcal{C}_y(\bm\theta^{*}_{x})^\top \left( \tilde{\bm\theta}^{*}_{y} - \bm\theta^{*}_{x} \right)\nonumber\\
\Leftrightarrow\;&\mathcal{C}_y(\bm\theta^{*}_{x}) \approx - \nabla_{\bm\theta} \mathcal{C}_y(\bm\theta^{*}_{x})^\top \left( \tilde{\bm\theta}^{*}_{y} - \bm\theta^{*}_{x} \right).
\end{align}

Since this is a single scalar equation in the vector unknown $\tilde{\bm\theta}^{*}_{y}$, we determine the minimum-norm solution by using the Moore-Penrose pseudo-inverse of the gradient \cite{courrieu2008fastcomputationmoorepenroseinverse}. Specifically, the pseudo-inverse of $\nabla_{\bm\theta} \mathcal{C}_y(\bm\theta^{*}_{x})$ is given by:

\begin{align}
\left[\nabla_{\bm\theta} \mathcal{C}_y(\bm\theta^{*}_{x})\right]^+ = \frac{\nabla_{\bm\theta} \mathcal{C}_y(\bm\theta^{*}_{x})^\top}{\|\nabla_{\bm\theta} \mathcal{C}_y(\bm\theta^{*}_{x})\|^2}.
\end{align}

Consequently, the corresponding parameters for $y$ are estimated:

\begin{align}
\tilde{\bm\theta}^{*}_{y} \approx \bm\theta^{*}_{x} - \frac{\nabla_{\bm\theta} \mathcal{C}_y(\bm\theta^{*}_{x})}{\|\nabla_{\bm\theta} \mathcal{C}_y(\bm\theta^{*}_{x})\|^2}\, \mathcal{C}_y(\bm\theta^{*}_{x}).
\label{eq:approx}
\end{align}

notated as $\tilde{\bm\theta}^{*}_{y}=\tilde{\frak{E}}_{\bm\theta_{x} \to \bm\theta_{y}}(\bm\theta^{*}_{x})$. The Equation~\eqref{eq:approx} provides an efficient means to adjust $\bm\theta^{*}_x$ for the small change from $x$ to $y$ without running $\verb|compilation|_U(\ldots)$ for ${\bm\theta}^{*}_{y}$ from $\mathbbm{1}$. If $\mathcal{C}_y(\tilde{\bm\theta}^{*}_y)<\tau$, $\bm\theta^{*}_{y}\equiv\tilde{\bm\theta}^{*}_y$, otherwise, $\bm\theta^{*}_{y}\gets\verb|compilation|_{U}(\ldots)$ with $\bm\theta^{(0)}=\tilde{\bm\theta}^{*}_y$. Evaluating \eqref{eq:approx} take $2m$ and $1$ quantum evaluation for $\nabla_{\bm\theta}\mathcal{C}_y(\bm\theta^*_x)$ and $\mathcal{C}_y(\bm\theta^{*}_{x})$, respectively, using the same resource for one compilation's iteration. However, if $\mathcal{C}_y(\bm\theta^{*}_{x})<\tau$, then $\bm\theta^{*}_{y}=\bm\theta^{*}_{x}$; there is no need for evaluating $\nabla_{\bm\theta}\mathcal{C}_y(\bm\theta^*_x)$.

\subsection{Fast compilation}

In standard quantum compilation, we encode $|\psi\rangle$ as fixed unitary $V$. In practice, $|\psi\rangle=V|0\rangle^{\otimes n}$ is prepared using the amplitude encoding algorithm \cite{10.5555/3511065.3511068}. However, this method is not scalable, as it requires circuit depths of $\mathcal{O}(2^{2n}) $. The unitary $U(\bm\theta)V^{\dagger}(x)$ is then transpiled on the quantum simulators (state-vector, tensor-network, decision diagram, etc.) or real quantum hardware (e.g., Vigo, Montreal, Quito) and conduct the measurement. 
Consequently, the Equation~\ref{eq:optimize} enables compression by reducing the complexity of a large circuit into a more efficient, smaller circuit. Because $\verb|compilation|_U(\ldots)$ will be called $NM$ times, we compute the term $d(\psi^{(t)}, \psi)$ directly as the inner product between two arrays instead of $\mathcal{M}_{\mathbb{Z}_0}(|\langle \bm 0|V^{\dagger}(x)U(\bm\theta^{(t)})|\bm 0\rangle|^2)$. This method reduces the execution time for $NM \times n_{\text{iter}}$ quantum evaluations; note that it is only valid in case we simulate measurement operators on a classical computer.

The structure of $U$ is also optimized; rather than using hypergraphs with multi-CZ gates which require much simulation time; we use $W_\text{chain}+ZXZ$. This ansatz is separated as ($L$ times) entangled and rotation parts; these parts are made from $n$ $CX$ and $3n$ $R_j\;(j\in\{x,y,z\})$ which are the basic gates. The number of parameters and circuit depth of $n$-qubit $W_\text{chain}+ZXZ$ are $3nL$ and $6L$, respectively.

\section{Quantum image encryption application}

Image encryption relates closely to the image exchange process. In the symmetric encryption and decryption processes, the sender encrypts the image with a secret key $pk$, and the receiver must use the same key to decode and receive the original image. If the image is encrypted without suffering/compression, the image patterns will not be hidden, and the attacker can obtain valuable information without decrypting. To cover this problem, the user can use QIC to compress the image as $\{\bm\theta^*\}$ then suffer and encrypt it; this method is similar to the neural network approach \cite{wang2022new}. The neural network approach utilizes deep learning models to learn the encryption process from the pairs $\{\text{original image}, \text{encrypted image}\}$. Then, the trained weight is only shared between the sender and receiver. This approach incurs additional costs for the training and testing process than the traditional approaches, such as steganography-based \cite{yildirim2021steganography} 
and DNA-based \cite{durafe2022image}. The QIC-based also requires cost for $NM$ $\verb|compilation|_U(\ldots)$.

\section{Experiments}
\label{sec:experiments}

\begin{figure*}[ht]
    \centering
    \includegraphics[width=0.99\linewidth]{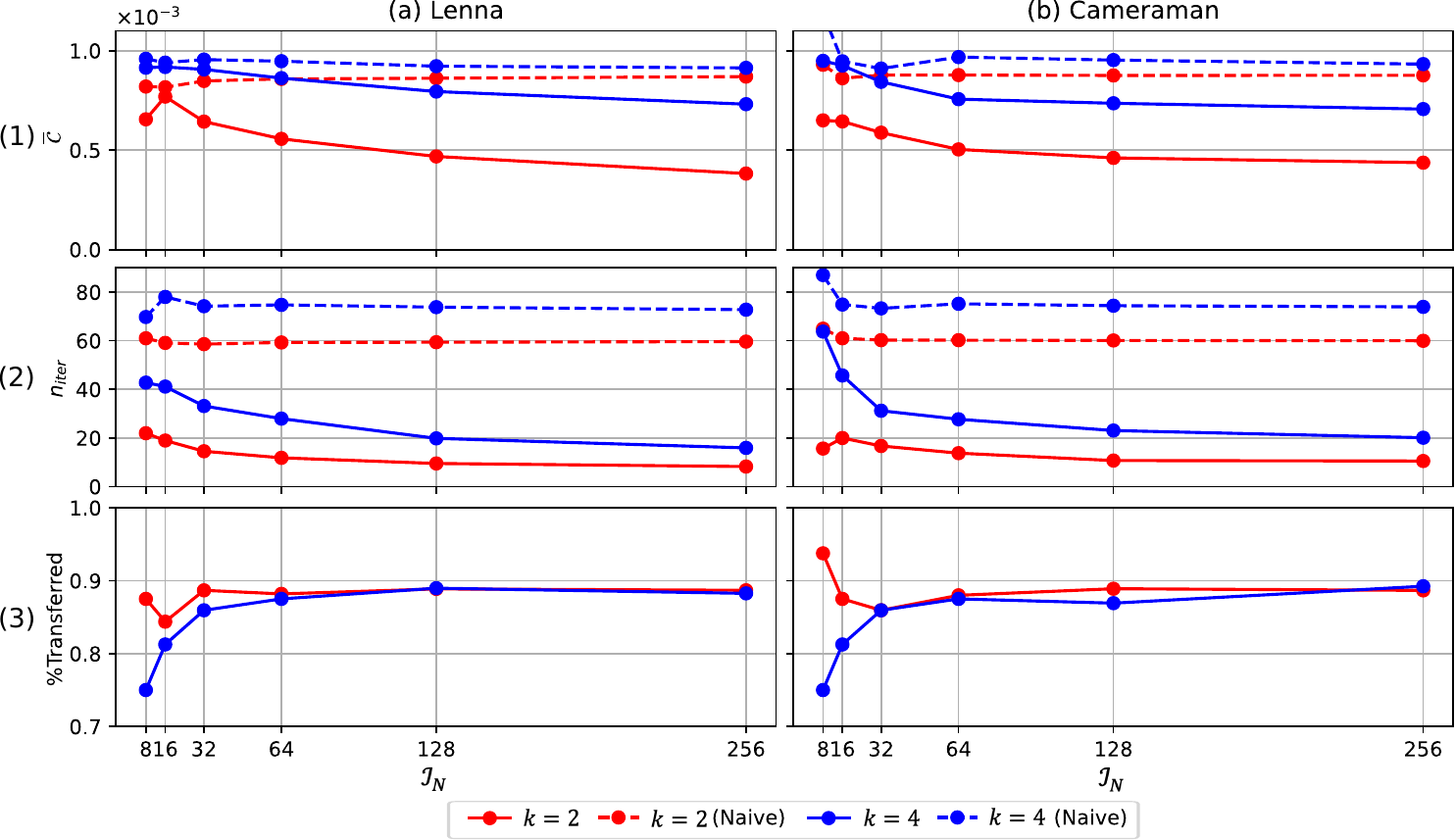}
    \vspace{-0.2cm}
    \caption{Experiment's result on (a) Lenna image and (b) Cameraman image. The investigated properties include (1) average cost value, (2) average minimal number of iterations, and (3) ``Transferred'' percent. We experimented with two images: Lenna and Cameraman. The original size of these images is $256\times256$ which is scaled as $8\times8, 16\times 16,\ldots$ and $128\times 128$. The fast method and the naive method's result are presented as normal (----) and dotted lines (\hdashrule[0.5ex]{0.4cm}{0.5pt}{0.5mm}), respectively.}
    \label{fig:results}
\end{figure*}

The algorithm is implemented in the Python 3.11 version with Pennylane as the core library \cite{bergholm2022pennylaneautomaticdifferentiationhybrid}. Our proposed algorithm is compared with the previous version \cite{hai_encrypted} (as a naive method) in case of compression loss $\overline{\mathcal{C}}=\sum_{i=1}^{N}(\sum_{j=1}^{M}\mathcal{C}^*_{i,j})$ and the average minimal number of iterations over all blocks $\{\mathcal{B}_{i,j}\}$. The final metric $\%\text{Transferred}$, is calculated as the number of ``estimate'' blocks divided by the number of blocks. The experiment setting is summarized as Table~\ref{tab:property}.

\begin{table}[ht]
    \centering
    \resizebox{0.49\textwidth}{!}{
    \begin{tabular}{|c|c|c|c|c|c|}
        \hline
        \textbf{Property} & $n_{\text{iter}}$ & $\tau$ & $\alpha$ (learning rate) & $L$ & $R$\\ \hline 
        \textbf{Value} & 100 & $10^{-3}$ & $0.1$ & $n$ & $1$\\ \hline
    \end{tabular}}
    \vspace{0.2cm}
    \caption{The properties setting for all trials}
    \label{tab:property}
\end{table}
As shown in Fig.~\ref{fig:results}, our proposed algorithm not only requires a smaller number of iterations but also achieves lower loss values. A good sign is that these two metrics tend to decrease based on the image size. The naive method keeps them stabilized because the naive method just simply repeats the same computation on more blocks. That means the fast QIC operates more efficiently on large images. For example, with the Lenna image and $k=2$, the proposal got the loss from $0.6\times10^{-3}$ (versus $0.8\times10^{-3}$) at $\mathcal{I}_N=8$ to $0.38\times 10^{-3}$ (versus $0.87\times 10^{-3}$) at $\mathcal{I}_N=256$. The average minimal number of iterations from the fast QIC is less than $\bm{64\%}$ ($21.94$ versus $61.00$) at $\mathcal{I}_N=8$ and less than $\bm{86\%}$ ($8.22$ versus $59.63$) at $\mathcal{I}_N=256$. The same phenomenon happens on $k=4$ and Cameraman's image. The noticed property is the ``transferred'' percent in Fig.~\ref{fig:results}-(3) which tends to coverage at $\textbf{88\%}$ for large image sizes.

\section{Conclusion}
\label{sec:conclusion}

By taking advantage of a hybrid quantum-classical model, the QIC provides a different level of compression based on the number of qubits ($n$) but still maintains a low loss. Meanwhile, all traditional methods must strike a balance between the compressed size and image quality. However, the weakness of QIC is the long runtime for $NM$ compilation for each image, making it infeasible for real scenarios. The Fast QIC, presented in this research, tries to optimize these calling functions by considering adjacency blocks first. This approach proves the large resources and loss reduction, especially better on higher-resolution images. As quantum computing technology progresses, the potential for further advancements in image processing and compression methodologies is increasingly promising. 

\section*{Code availability}

The code is available at \url{https://github.com/vutuanhai237/QuantumCompressor}.

\section*{Acknowledgment}

This work is supported by the VNUHCM-University of Information Technology’s Scientific Research Support Fund under Grant No. D1-2023-48.


\bibliographystyle{IEEEtran}
\bibliography{ref.bib}

\end{document}